\documentstyle[12pt,psfig]{article}
\catcode`\@=11
\def\marginnote#1{}
\newcount\hour
\newcount\minute
\newtoks\amorpm
\hour=\time\divide\hour by60
\minute=\time{\multiply\hour by60 \global\advance\minute by-
\hour}
\edef\standardtime{{\ifnum\hour<12 \global\amorpm={am}%
    \else\global\amorpm={pm}\advance\hour by-12 \fi
    \ifnum\hour=0 \hour=12 \fi
    \number\hour:\ifnum\minute<100\fi\number\minute\the\amorpm}}
\edef\militarytime{\number\hour:\ifnum\minute<100\fi\number\minute}
\def\draftlabel#1{{\@bsphack\if@filesw {\let\thepage\relax
  \xdef\@gtempa{\write\@auxout{\string
    \newlabel{#1}{{\@currentlabel}{\thepage}}}}}\@gtempa
    \if@nobreak \ifvmode\nobreak\fi\fi\fi\@esphack}
     \gdef\@eqnlabel{#1}}
\def\@eqnlabel{}
\def\@vacuum{}
\def\draftmarginnote#1{\marginpar{\raggedright\scriptsize\tt#1}}
\def\draft{\oddsidemargin -.5truein
        \def\@oddfoot{\sl preliminary draft \hfil
        \rm\thepage\hfil\sl\today\quad\militarytime}
        \let\@evenfoot\@oddfoot \overfullrule 3pt
        \let\label=\draftlabel
        \let\marginnote=\draftmarginnote

\def\@eqnnum{(\theequation)\rlap{\kern\marginparsep\tt\@eqnlabel}%
\global\let\@eqnlabel\@vacuum}  }
\def\preprint{\twocolumn\sloppy\flushbottom\parindent 1em
        \leftmargini 2em\leftmarginv .5em\leftmarginvi .5em
        \oddsidemargin -.5in    \evensidemargin -.5in
        \columnsep 15mm \footheight 0pt
        \textwidth 250mmin      \topmargin  -.4in
        \headheight 12pt \topskip .4in
        \textheight 175mm
        \footskip 0pt

\def\@oddhead{\thepage\hfil\addtocounter{page}{1}\thepage}
        \let\@evenhead\@oddhead \def\@oddfoot{} \def\@evenfoot{}
}
\def\titlepage{\@restonecolfalse\if@twocolumn\@restonecoltrue\onecolumn
     \else \newpage \fi \thispagestyle{empty}\c@page\z@
        \def\thefootnote{\fnsymbol{footnote}} }
\def\endtitlepage{\if@restonecol\twocolumn \else  \fi
        \def\thefootnote{\arabic{footnote}}
        \setcounter{footnote}{0}}  
\catcode`@=12
\relax
\def\be{\begin{equation}}
\def\ee{\end{equation}}
\def\bea{\begin{eqnarray}}
\def\eea{\end{eqnarray}}
\def\simlt{\stackrel{<}{{}_\sim}}
\def\simgt{\stackrel{>}{{}_\sim}}

\def\NPB#1#2#3{{\it Nucl.~Phys.} {\bf{B#1}} (19#2) #3}
\def\PLB#1#2#3{{\it Phys.~Lett.} {\bf{B#1}} (19#2) #3}
\def\PRD#1#2#3{{\it Phys.~Rev.} {\bf{D#1}} (19#2) #3}
\def\PRL#1#2#3{{\it Phys.~Rev.~Lett.} {\bf{#1}} (19#2) #3}

\relax

\begin{document}
\topmargin-1.8cm
%
\begin{titlepage}
\begin{flushright}
CERN-TH/95-261\\
\end{flushright}
\begin{center}{\Large\bf
Physics at LEP and Yukawa Coupling Unification}\footnote{
Talk presented at {\bf SUSY95}, Palaiseau, France,
May 1995. To appear in the Proceedings.} \\
\vspace*{3.0ex}
\vskip 0.7 cm
{\bf  C.E.M. Wagner}\\
\vspace*{3.0ex}
\vskip 0.7 cm
CERN, TH Division, CH--1211 Geneva 23, Switzerland \\
\end{center}
\vspace*{2.5ex}
\vskip 3.5 cm
\begin{center}
{\bf Abstract}
\end{center}
\vspace*{3.ex}
\begin{quote}
We discuss the impact of the recent precision measurements at LEP
on the minimal supersymmetric standard model.
We show that the values of $\tan\beta$ necessary to induce large
positive corrections to $R_b$  are the same as the ones preferred by
the unification of the bottom and $\tau$
Yukawa couplings in the MSSM for the current
measured value of $M_t$. We discuss the relevance
of the preferred parameter space for the Higgs and sparticle
searches at LEP2. Remarkably, it follows that
the LEP measurements can provide
information about the structure of  soft supersymmetry breaking
parameters at $M_{GUT}$. Finally, we  briefly discuss the properties of
a supersymmetric model with four generations,
for which the fit to $R_b$ naturally
improves with respect to the one in the Standard Model.
\end{quote}
\vskip2.cm
\begin{flushleft}
CERN-TH/95-261\\
October 1995 \\
\end{flushleft}
\end{titlepage}
\setcounter{footnote}{0}
\setcounter{page}{0}
\newpage
One of the strongest motivations for the
analysis of grand unified scenarios, which lead to the
minimal supersymmetric standard model (MSSM) as
an effective low-energy theory, comes from
the successful prediction for the weak mixing
angle  within this framework \cite{SUSYG}.
Apart from yielding
relations between the gauge couplings, most grand unified
scenarios relate the values of the third-generation
Yukawa couplings at $M_{GUT}$,
leading to definite low-energy predictions for these
quantities.
The recent measurements of the top quark mass \cite{CDF}
at the Tevatron  collider allow a restriction of the parameter space
preferred by bottom-$\tau$
Yukawa coupling unification to two distinct regions:
the region of very low $\tan\beta$, associated with the infrared
fixed point for the top quark mass, and the region of very large
$\tan\beta \simeq m_t/m_b$ \cite{Ramond,Yukawa,largetb}.
It is the purpose of this talk,
in the light of the most recent precision measurements at LEP,
to investigate the
most relevant experimental signatures associated with
the theoretical scenario of Yukawa coupling unification.

The electroweak precision measurements at LEP and SLD are in
remarkable agreement with the predictions of the Standard Model,
in the presence of a heavy top quark, $M_t \simeq 175$~GeV \cite{sumfit}.
The most recent experimental data show, however,
that the partial quark width $R_b$ ($R_c$)
[$R_q = \Gamma(Z \rightarrow q \bar{q})/\Gamma(Z \rightarrow
{\rm hadrons})$],  lies more than three (two) standard deviations
above (below) the standard model fit for these quantities.
If these measurements
were taken at face value, they would point towards the presence
of new physics at the weak scale,
which can lead to large $Zq\bar{q}$ vertex corrections.
{}From the experimental point of view, however, the measurement of
$R_c$ is subject to large experimental uncertainties.
Moreover, if new
physics were responsible for the deviation in $R_c$ it should
affect also the total light quark widths in a very precise way
not to spoil the predictions for the total hadron width.
The measurement of $R_b$ is instead more reliable; hence, the present
discrepancy with the standard model prediction deserves further
investigation.
The  LEP electroweak working group
provides also a determination of $R_b$ based on the assumption that $R_c$ is
approximately given by its standard model value:
$R_b \; = \; 0.2205 \pm 0.0016$, a value which is still more than
three $\sigma$ above the standard model prediction for this quantity:
$R_b^{SM} \simeq 0.2156$.

Within the minimal supersymmetric standard model, large
corrections to $R_b$ are always associated with regions
of large top and/or bottom Yukawa couplings \cite{BF,Gordy}.
These couplings
appear in the chargino-stop (charged Higgs-top) and
neutralino-sbottom (neutral Higgs-bottom) one-loop
contributions to the $Zb\bar{b}$ vertex. The value of
the top Yukawa coupling is maximized for low values of
$\tan\beta$, and hence  large corrections to $R_b$
are observed close to the top quark mass infrared
fixed-point solution \cite{CW}. Since the charged Higgs contributions
reduce the value of $R_b$, large positive corrections may
only be obtained for large values of the charged Higgs mass.
Stop-chargino loops give, instead, positive contributions
to $R_b$; thus low values for the stop mass and, most important,
for the chargino mass, within the reach of LEP2,
are preferred. Large bottom Yukawa couplings, instead,
may only be obtained for large values of $\tan\beta
\simeq m_t/m_b$. In this region of parameters, the total
Higgs contributions become positive for $m_A \simlt 70$ GeV,
and for sufficiently low values of $m_A$ they are larger
than the genuine supersymmetric ones \cite{Gordy,Joan}.
Low values of the third-generation  squark,
chargino and neutralino masses are also helpful in order to get
large radiative corrections~to~$R_b$.

It is remarkable that the regions of $\tan\beta$ preferred by
the best fit to precision measurement data
and by the requirement of bottom-$\tau$
Yukawa unification coincide. Let us first concentrate in the
low $\tan\beta$ region, associated with the infrared fixed point
of the top quark mass.
%
One of the interesting  properties of the top quark  mass
infrared fixed-point solution
is that it implies also a fixed-point solution
for some soft supersymmetry-breaking parameters \cite{COPW}.
Summarizing the results for the relevant
low-energy mass parameters at the infrared fixed point
we have \cite{CW}:
\begin{eqnarray}
\label{eq:massp}
m_{H_2}^2 & \simeq &  \frac{m_{H_2}^2(0)}{2} - m_Q^2(0)
- 3.5 M_{1/2}^2
\;\;\;\;\;\;\;\;\;\;\;\;\;\;
A_t \simeq -2 M_{1/2}
\\
m_Q^2 & \simeq & \frac{2 m_Q^2(0)}{3} - \frac{m_{H_2}^2(0)}{6}
  + 6 M_{1/2}^2
\;\;\;\;\;\;\;\;\;\;\;\;\;\;
m_U^2  \simeq  \frac{m_Q^2(0)}{3} - \frac{m_{H_2}^2(0)}{3}
+ 4 M_{1/2}^2
\nonumber\\
\mu^2 & \simeq & \left[
m_{H_1}^2(0) +  \left(\frac{2 m_Q^2(0) - m_{H_2}^2(0)}{2}
\right) \tan^2 \beta
 +
M_{1/2}^2 \left( 0.5 + 3.5 \tan^2\beta \right) \right] \frac{1}{\tan^2\beta
  - 1 }
\nonumber
\end{eqnarray}
where $m_i^2(0)$ and $M_{1/2}$
denote the values of the scalar mass parameters
and the gaugino masses at $M_{GUT}$ (for notation and sign conventions,
see Ref. \cite{CW}). Observe that $A_t$ and $M_U^2 = m_Q^2 + m_U^2
+ m_{H_2}^2$ are  determined
by the parameters of the gauge sector of the theory.

Another  very important feature  of the spectrum at the infrared fixed point
is associated with the Higgs sector.
The Higgs spectrum consists of  three neutral scalar states --two
CP-even, h and H,  and  one CP-odd, A--  and two charged scalar states
H$^{\pm}$. The mass of the CP-odd scalar Higgs is approximately given by
its tree level expression, $m_A^2 \simeq m_1^2 + m_2^2$,
\bea
m_A^2
&  \simeq &
 \left[ m_{H_1}^2(0) + \left(
\frac{ 2 m_Q^2(0) -  m_{H_2}^2(0)}{2} \right)
 + 4 M_{1/2}^2 \right] \frac{\tan^2 \beta + 1}{\tan^2 \beta -1}
\label{eq:mA}
\eea
{}From Eq. (\ref{eq:mA}) it follows that, for lower values of $\tan \beta$,
the value  of the CP-odd mass eigenstate  is enhanced.
Moreover,
larger values of $m_A$ imply as well that the charged Higgs and
the heaviest CP-even Higgs  become  heavy in such a regime.
For these large values of $m_A$,
$m_h$ acquires values close to its upper bound, which is independent
of the exact value of the CP-odd mass.
Furthermore, for a given value of the
physical top quark mass, the infrared fixed-point solution  is associated
with the minimum value of $\tan \beta$ compatible  with the perturbative
consistency of the theory. For  $\tan \beta\geq 1$, lower values
of $\tan \beta$ correspond to lower values of the tree-level lightest
CP-even mass,
$m_h^{tree} = M_Z |\cos 2 \beta|$, and
after the inclusion of the radiative corrections,
the fixed point still gives the lowest
possible value of $m_h$ for any given value
of $M_t$ \cite{BABE,Cartalk}.
Indeed, taking into account the most relevant
two-loop corrections \cite{HHo},
an approximate expression for the Higgs mass
may be obtained \cite{CEQW},
\begin{eqnarray}
m_h^2& = & M_Z^2\cos^2 2\beta\left( 1-\frac{3}{8\pi^2}\frac{m_t^2}
{v^2}\ t\right) \nonumber \\
\label{mhsm}
& + & \frac{3}{4\pi^2}\frac{m_t^4}{v^2}\left[ \frac{1}{2}\tilde{X}_t + t
+\frac{1}{16\pi^2}\left(\frac{3}{2}\frac{m_t^2}{v^2}-32\pi\alpha_3(M_t)
\right)\left(\tilde{X}_t t+t^2\right) \right]
\end{eqnarray}
where $\tilde{X}_{t}  = 2 \tilde{A}_t^2/M_{\rm SUSY}^2
                  \left(1 - \tilde{A}_t^2/12 M_{\rm SUSY}^2 \right)$,
$\tilde{A}_t  =  A_t-\mu\cot\beta$,
$t=\log\frac{ M_{\rm SUSY}^2}{M_t^2}$, $m_t = m_t(M_t)$
and $v = 174$ GeV.
Taking the value of $\tilde{X}_t$ which maximizes the Higgs mass,
and for $M_{SUSY} \simlt 1$~TeV
and $M_t \simlt 175$ GeV, an upper bound $m_h \simlt 100$
GeV is obtained. The most recent experimental analyses have shown that
for such range a of values of $M_t$ the infrared fixed point solution can
be fully tested at LEP2~\cite{interim}.

A question that immediately
arises is  that of the maximal value of $R_b$,
which may be obtained for arbitrary choices of the
boundary conditions for the mass parameters.
The larger  variations of $R_b$ with respect to its
Standard Model value are found for  solutions such that
the  right-handed component of the lightest stop is maximized by
requiring low (large) values for the mass parameter $m_U$ ($m_Q$).
These conditions also imply that the stop contributions
to the variable $\Delta \rho$ are small.
Moreover, the lightest
stop and chargino masses and the $\mu$ parameter
should acquire values below $M_Z$.
Finally, the charged Higgs must be heavy, $m_{H^+} \gg M_Z$.
Light stops demand significant
mixing effects, $A_t \simeq m_Q$.  Taking $M_t \simeq$ 160--180 GeV
($\tan\beta \simeq $ 1.1--1.6) and imposing the
experimental constraints coming from
$\Delta \rho$ and the branching ratio $b \rightarrow s \gamma$, the
maximum value of $R_b$ achievable is  $R_b \simeq 0.2180$.
This value of $R_b$ has the effect of changing the strong gauge
coupling determination at LEP,
coming from $R_\ell = \Gamma_\ell/\Gamma_h$,
where $\Gamma_\ell$ is the $Z$ leptonic width, to a value
$\alpha_3(M_Z) \simeq 0.116$.

Having found these solutions, it is important to analyse for
which particular values of the boundary conditions for
the scalar mass parameters they are  obtained.
The set of values \cite{CW}
\begin{eqnarray}
m_Q^2(0) = m_U^2(0) \simeq 0; \;\;\;\;\;\;\;\;\;
m_{H_2}^2(0) \simeq 12 M_{1/2}^2;
\;\;\;\;\;\;\;\;\;
m_{H_1}^2(0) &\simeq& 2.5 \;\tan^2\beta \; M_{1/2}^2,
\label{eq:opcond}
\end{eqnarray}
with $M_{1/2} > 300$ GeV, leads to solutions in the desired range.
The first two relations in Eq. (\ref{eq:opcond})
are necessary in order to
minimize the value of $m_U$ while keeping $A_t \simeq m_Q$.
The last relation is required in order to minimize the value
of the $\mu$ parameter.  Interestingly,
the solutions that maximize $R_b$
demand a small value of the  squark mass parameters at the grand
unification scale.
Finally, the charged Higgs mass is naturally large
within this scheme. \\

We shall now discuss the large $\tan\beta$ region \cite{we5},
for which,
as explained above, low values of the CP-odd
Higgs mass, $m_A \simlt 70$ GeV,  are preferred
to improve the agreement with the value of $R_b$ measured at LEP.
An  important tree level relation is obtained for large values of
the ratio of Higgs vacuum expectation values, $\tan\beta > 10$, for
which the tree level value of the lightest CP-even
Higgs mass is equal to $M_Z$, whenever the
CP-odd Higgs mass $m_A$ is larger than $M_Z$, while for $m_A \leq
M_Z$, $m_h = m_A$.  This tree level relation is approximately stable
under radiative corrections, with the only difference  that the
range for which the equality $m_h = m_A$ holds, extends to values
of $m_A$ somewhat larger than $M_Z$. Therefore, large values of
$\tan\beta$ and values of the CP-odd Higgs in the desired range imply
\begin{equation}
m_h < M_Z,
\end{equation}
and hence the neutral Higgs sector will be tested at LEP2
within this framework.

Moreover, the charged Higgs mass is determined through the
CP-odd Higgs mass value, $m_{H^{\pm}}^2 \simeq m_A^2 + M_W^2$.
For  $m_{H^{\pm}}
\simlt 130$ GeV, the branching ratio
${\rm BR}(b \rightarrow s \gamma)$ becomes larger than the
presently allowed experimental values, unless the
supersymmetric particle contributions suppress the
charged Higgs enhancement of the decay rate.
The most important supersymmetric
contributions to this rare bottom decay
come from the chargino-stop one-loop diagram \cite{bsga}.
The chargino contribution to the $b \rightarrow s \gamma$
decay amplitude depends on the soft supersymmetry breaking
mass parameter $A_t$ and on the supersymmetric mass parameter
$\mu$, for very large values of $\tan\beta$, it is
given by
\be
A_{\tilde{\chi}^+} \simeq \frac{m_t^2}{m_{\tilde{t}}^2}
\frac{A_t \mu}{m_{\tilde{t}}^2} \tan\beta \; G\left(
\frac{m_{\tilde{t}}^2}{\mu^2}\right),
\ee
where $G(x)$ is a function that takes values of order 1
when the characteristic stop mass $m_{\tilde{t}}$
is of order $\mu$ and grows for low values
of $\mu$. One can show that for positive (negative)
values of $A_t \times \mu$ the chargino contributions are of
the same (opposite) sign as the charged Higgs ones. Hence,
to get an acceptable $b \rightarrow s \gamma$
decay rate, negative values for $A_t \times \mu$ are required.
Considering the running of the soft supersymmetry-breaking
terms from high energies,
the value of $A_t$ is approximately given by
$A_t = A_0 \left( 1 - \frac{Y_t}{Y_f} \right) - M_{1/2}
\left(4 - 2 \frac{Y_t}{Y_f} \right)$ \cite{wefour},
where $A_0$ and
$M_{1/2}$ are the values of the trilinear coupling and
the common gaugino mass at the unification
scale and $Y_f$ is the fixed-point value for $Y_t$. For values
of the top quark mass $M_t \simgt 160$ GeV,
the mass parameter $A_t$ is opposite in sign to $M_{1/2}$,
unless $A_0$ is more than one order
of magnitude larger than $M_{1/2}$.
Hence, in order to get acceptable values for the branching
ratio ${\rm BR}(b \rightarrow s \gamma)$,
\begin{equation}
\mu \times M_{1/2} > 0.
\end{equation}

Large radiative corrections to the bottom mass arise
from the effective coupling of the bottom quark to the
Higgs $H_2$
at the one loop level \cite{Hall,wefour}. Indeed,
although the effective coupling of the down
quarks to $H_2$   is small
in comparison to the down quark Yukawa coupling to the Higgs $H_1$,
for large values of $\tan\beta$ the vacuum expectation value
of $H_2$ is much larger than that  of $H_1$,
and hence large down quark mass
corrections may be obtained through this effect \cite{downc}.
Quantitatively,
$\Delta m_b/m_b \simeq \tan\beta \times
2 \alpha_3 M_{\tilde{g}} \mu/(3 \pi m_{\tilde{b}}^2)$,
where $Y_t = h_t^2/(4\pi)$  and
$m_{\tilde{b}}^2$ is the value of the heaviest
sbottom mass (for more precise expressions
see Refs. \cite{Hall,wefour,downc}).
Due to the constraints on the parameters $\mu$ and $M_{1/2}$
discussed above, the bottom mass
corrections are positive within this scheme. Positive mass corrections
mean that, in order to get an acceptable values
for the physical bottom mass, the value of the predicted bottom mass
before addition of the finite corrections, $\tilde{M}_b$,
should be given by  $\tilde{M}_b \leq 5.2 {\rm GeV}$.
This puts strong constraints on the allowed values of $M_t$. For
$\alpha_3(M_Z) \simgt 0.125,0.120,0.115$, the top quark mass
$M_t \simgt 180,170,160$ GeV (See Ref. \cite{wefour}). \\

Finally, let us discuss the model with four generations \cite{fourgen}:
If one considers
the $R_b$ measurement alone within the context of the standard model,
one would conclude that $m_t \simlt m_W$.
Previous top quark searches at hadron
colliders are able to close the window between
the top mass lower bound coming from the $W^\pm$ width, 62 GeV, and
$m_W + m_b \simeq 85$ GeV,
assuming that $t\rightarrow bW^\star$ is the dominant top-quark decay mode.
However, if the top quark had any two-body decay modes (due
to new physics processes), and if these modes rarely produced
leptons, then a top quark in this mass region would not have been
detected in any experiment.
A  well motivated and
experimentally acceptable scenario occurs in supersymmetric models
in which the decay
$t\rightarrow\widetilde t\widetilde\chi^0_1$ is kinematically allowed
(where $\widetilde t$ is the lightest top squark and
$\widetilde\chi^0_1$ is the lightest neutralino \cite{rudaz}).
We choose $m_t\simeq m_W$, $M_{\tilde t}\simeq 50$~GeV and
$M_{\tilde\chi^0_1}\simeq 25$~GeV.  Then, the dominant decay chain is
$t\rightarrow\widetilde t\widetilde\chi^0_1$ followed by
$\widetilde t\rightarrow c\widetilde\chi^0_1$ through a one-loop
process,  which rarely produces a hard isolated lepton.

The fourth-generation top quark decays
$t^\prime\rightarrow bW^+$ can be the source of the CDF and D0 events.
For this, the $t^\prime$--$b$ mixing angle
($V_{t^\prime b}$) must not be too small; otherwise, the
$t^\prime \rightarrow b^\prime + W^+$ decay mode
will dominate.   For definiteness, we
choose $|V_{t^\prime b}/V_{t^\prime b^\prime}|=0.1$.
It is straightforward to show that, for $m_{t^{\prime}}
\simeq 175$ GeV,
a ${\rm BR}(t^{\prime}
\rightarrow b W^+)$ close to 1 is obtained if $m_{b^{\prime}}
\simgt 105$~GeV.
Moreover, imposing the  requirement that the Yukawa
couplings remain perturbative up to the grand unification
scale, we obtain that two properties must be fulfilled: first,
the $b^\prime$ mass must be close to its lower bound,
$m_{b^{\prime}} \simeq 105$ GeV.
Second, the $b^\prime$ and $t^\prime$ Yukawa couplings must lie close to
their fixed-point values and $\tan\beta \simeq 1.6$.

The fourth-generation lepton Yukawa couplings are also
strongly constrained by the experimental bound on the
lepton masses.  The
bound on the $\tau^\prime$ lepton mass implies that its
Yukawa coupling must get large values at the grand unification
scale. Assuming that the fourth-generation neutrino ($N$) is a Dirac fermion,
one can also add the requirement that its  Yukawa coupling takes large
values at $M_{GUT}$. The resulting lepton
masses are: $m_{\tau^\prime}\simeq 50$~GeV and $m_N
\simeq 80$~GeV. Remarkably, the fourth generation fermion
masses within this model are consistent with the unification of
the four fourth generation Yukawa couplings at the GUT scale.

The constraints coming from the oblique corrections to
the gauge bosons, most notably the $\Delta\rho$ parameter,
put strong constraints on the sparticle masses.
Indeed, since the top quark mass is less than half of its
standard value, the contribution of the $t$--$b$ doublet
to $\Delta\rho$ is reduced by a factor of 4.
Due to the large values of $m_{b^\prime}$,
the fourth-generation quark contribution
to $\Delta\rho$ is approximately the same as the
third-generation one. Hence, the supersymmetric particle virtual
effects must account for  roughly half of the total
contribution to  $\Delta\rho$.
One must maximize the off-diagonal squark mixing while keeping
the diagonal squark mass parameters as small as possible.  However,
the latter cannot be too small; otherwise the radiative corrections to
the light Higgs mass will be reduced leading to a value of $m_h$ below
the current LEP bound.

For $m_Q \simeq m_U$, relevant contributions to $\Delta \rho$ induced
by the third-generation squarks may only be obtained if $m_Q$ is
somewhat above the lower bounds on this quantity coming from the
present experimental bounds on the lightest sbottom mass, implying
large values of the stop mixing mass parameters.
Non-negligible mixing in the fourth-generation also enhances the
fourth generation squark contributions to $\Delta\rho$.  The maximum
effect is limited phenomenologically by
a lower bound on the mass of $\widetilde b^\prime$.
In order that $t^\prime\rightarrow bW^+$ remain the dominant decay, one
must kinematically forbid $t^\prime\rightarrow\widetilde b^\prime\widetilde
\chi_1^+$.
This in turn
determines the best value
for $\mu$, since $A_{t^\prime}$ and $A_{b^\prime}$ are determined by
their infrared fixed-point behaviour \cite{COPW,wefour},
$A_{t^\prime} \simeq A_{b^\prime} \simeq -200$ GeV.
Taking this into account we find that, for instance, if
$m_{Q} \simeq m_{U} \simeq$  275 GeV,
$\mu \simeq -420$ GeV
$m_{Q^\prime}  \simeq m_{{U}^\prime}  \simeq$
250 GeV, and $A_t \simeq 750$ GeV,
the contribution from the third- and fourth- generation squarks
is only slightly lower than that one of their fermion superpartners,
leading to acceptable values for $\Delta\rho$.
The price to pay is a very large
value of the stop mixing mass term
$\tilde{A}_t \simeq 1$ TeV. These large values of the mixing
mass parameter $\tilde{A}_t$ may render
the ordinary vacuum state  unstable \cite{CW,stable,stable2}.
Our model
predicts a light Higgs spectrum, $m_h \simeq 70$ GeV, and an
improved value of $R_b = 0.2184$, which is within $1.5 \sigma$
of the measured LEP value.
%
%

In summary, the three scenarios discussed above  are consistent
with Yukawa coupling unification. In each scenario the global fit
to the present electroweak precision measurement is significantly
improved if new particles, with  masses below the LEP2 kinematical limit,
are present: \\
a) Infrared fixed point: A chargino, two neutralinos
and a  stop. Furthermore, the lightest CP-even Higgs
(if $M_t \simlt 175$ GeV), must be observable at LEP2. \\
b) Large $\tan\beta$: The CP-odd and the lightest CP-even Higgs bosons,
a chargino, two neutralinos and a stop. \\
c) Model with four generations:  The lightest
CP-even Higgs, a charged and a neutral
lepton, a chargino, two neutralinos, a stop and the top quark. \\
{\bf Acknowledgement:} I would like to thank M. Carena, P. Chankowski,
H. Haber, M. Olechowski, S. Pokorski and M. Quir\'os for fruitful
and enjoyable collaborations.


\newpage
\begin{thebibliography}{99}
\bibitem{SUSYG}
S. Dimopoulos, S. Raby and F. Wilczek, \PRD{24}{81}{1681};\\
S. Dimopoulos and H. Georgi, \NPB{193}{81}{150};\\
L. Iba\~nez and G.G. Ross, \PLB{105}{81}{150}.
%
\bibitem{CDF} F.~Abe et al., CDF Collaboration, \PRD{50}{94}{2966};
\PRL{73}{94}{225}; preprint FERMILAB-PUB-95/022-E (2 March 1995) \\
S.~Abachi et al., D0 Collaboration,
\PRL{72}{94}{2138}; \PRL{74}{95}{2422}; preprint
FERMILAB-PUB-95/028-E (3 March 1995)
%
\bibitem{sumfit} P. Antilogus {\it et al.}
[LEP Electroweak Working Group], LEPEWWG/95-02 (1995).
%
\bibitem{Ramond} H. Arason, D. J. Casta\~no, B. Keszthelyi,
S. Mikaelian, E. J. Piard, P. Ramond and B. D. Wright,
Phys. Rev. Lett. 67 (1991) 2933;\\
S. Dimopoulos, L. Hall and S. Raby,
Phys. Rev. Lett. 68 (1992) 1984, Phys. Rev. D45 (1992) 4192.
\bibitem{Yukawa}  V. Barger, M.S. Berger and P. Ohmann,
\PRL{49}{94}{4908}; \\
P. Langacker and N. Polonsky, \PRD{47}{93}{4028}; {\bf D49} (1994) 1454; \\
M. Carena, S. Pokorski and C.E.M. Wagner,
\NPB{406}{93}{59}; \\
W.A. Bardeen, M. Carena, S. Pokorski and
C.E.M. Wagner, \PLB{320}{94}{110}.
%
\bibitem{largetb} M. Olechowski, S. Pokorski, \PLB{214}{88}{393};\\
B. Anantharayan, G. Lazarides and Q. Shafi, \PRD{44}{91}{1613};
%
\bibitem{BF} M. Boulware and D. Finnel, \PRD{44}{91}{2054}.
\bibitem{Gordy}
J.D. Wells, C. Kolda and G.L. Kane, \PLB{338}{94}{219};\\
P. Chankowski and S. Pokorski, Warsaw University
preprint, IFT-95/5, to appear in the Proceedings of \lq\lq Beyond the
Standard Model IV'', Lake Tahoe, CA, December 1994; MPI preprint
MPI-PhT/95-49 (1995); \\
A. Dabelstein, W. Hollik and W. M\"osle, Univ.
of Karlsruhe preprint, KA-THEP-5-1995 (1995).
%
\bibitem{CW} M. Carena and C.E.M. Wagner, Nucl. Phys. {\bf B452} (1995) 45.
%
\bibitem{Joan} D. Garcia and J. Sola, \PLB{354}{95}{335}.
%
\bibitem{COPW} M. Carena, M. Olechowski, S. Pokorski and C.E.M. Wagner,
\NPB{419}{94}{213}.
%
\bibitem{BABE} V. Barger, M.S.$\;$Berger and P. Ohmann,
\PRD{47}{93}{1093}; V. Barger, M.S. Berger,
P. Ohmann and R.J.N. Phillips, \PLB{314}{93}{351}.
%
\bibitem{Cartalk} C.E.M. Wagner, {\it Properties of SUSY particles},
eds. L. Cifarelli and V.A. Khoze (World Scientific, Singapore, 1993)
p. 469.
%
\bibitem{HHo} R.~Hempfling and A.H.~Hoang, \PLB{331}{94}{99}; \\
J.~Kodaira, Y.~Yasui and K.~Sasaki, \PRD{50}{94}{7035}; \\
J.A.~Casas, J.R.~Espinosa, M.~Quir\'os and A.~Riotto,
\NPB{436}{95}{3}; (E) {\bf B439} (1995) 466.
%
\bibitem{CEQW} M. Carena, J.R. Espinosa, M. Quiros and C.E.M. Wagner,
\PLB{355}{95}{209}; \\
M. Carena, M. Quiros and C.E.M. Wagner, CERN preprint CERN-TH/95-157,
August 1995, Submitted to {\it Nucl. Phys. B}.
%
\bibitem{interim} G. Altarelli et al., {\it Interim Report on the
Physics Motivations for an Energy Upgrade of LEP2},  CERN preprint
CERN-TH/95-151, CERN-PPE/95-78, June 1995.
%
\bibitem{we5} For a more extensive discussion, see
M. Carena, P. Chankowski, M. Olechowski, S. Pokorski
and C.E.M. Wagner, to appear.
%
\bibitem{wefour} M. Carena, M. Olechowski, S. Pokorski and
C.E.M. Wagner, \NPB{426}{94}{269}.
%
\bibitem{Hall} L.J. Hall, R. Rattazzi and U. Sarid,
\PRD{50}{94}{7048}; \\
R. Hempfling, \PRD{49}{94}{6168}.
%
\bibitem{downc} M. Carena, S. Dimopoulos, S. Raby and C.E.M. Wagner,
CERN preprint CERN-TH/95-53, hep-ph/9503488, to be published in Phys. Rev. D;\\
T. Bla\v{z}ek, S. Raby and S. Pokorski, Ohio Univ.
preprint OHSTPY-HEP-T-95-007, hep-ph/9504364, to be published in Phys. Rev. D.
%
\bibitem{stable} J.M. Fr\`ere, D.R.T. Jones and S. Raby,
\NPB{222}{83}{11};\\
L. Alvarez Gaum\'e, J. Polchinski and M. Wise, \NPB{221}{83}{495};\\
C. Kounnas, A. Lahanas, D.V. Nanopoulos and M. Quir\'os,
\NPB{236}{84}{438};\\
J.F. Gunion, H.E. Haber and M. Sher, \NPB{306}{88}{1}.
%
\bibitem{stable2}
P. Langacker and N. Polonsky, \PRD{50}{94}{2199};\\
J.A. Casas, A. Lleyda and C. Mu\~noz, preprint FTUAM 95/11 (1995).
%
\bibitem{bsgexp}
B. Barish et al., CLEO Collaboration, preprint CLEO CONF 94-1,
to appear in the Proceedings of the ICHEP94 Conference,
Glasgow, Scotland, July 1994.
%
\bibitem{bsga} S. Bertolini, F. Borzumati, A. Masiero and
G. Ridolfi, \NPB{353}{91}{591};\\
R. Barbieri and G. Giudice, \PLB{309}{93}{86}.
%
\bibitem{fourgen} For a more extensive discussion, see
M. Carena, H.E. Haber and C.E.M. Wagner, CERN preprint
CERN-TH/95-235, October 1995.
%
\bibitem{rudaz}
I.I. Bigi and S. Rudaz, \PLB {153}{85}{335}.
\end{thebibliography}
\end{document}